\documentclass{elsart}
\usepackage{psfig}

\begin{document}

\begin{frontmatter}

\title{$\eta$- and  $\eta^{\prime}$-meson production in the reaction 
  $p n \to d M$ near threshold}

\author[INR]{V.Yu. Grishina\thanksref{DFG}},
\author[ITEP]{L.A. Kondratyuk\thanksref{DFG}},
\author[Juelich]{M. B\"uscher\thanksref{DFG}},
\author[Juelich]{J. Haidenbauer},
\author[Seattle]{C. Hanhart\thanksref{hum}}, and
\author[Juelich]{J. Speth}
\address[INR]{Institute for Nuclear Research, 60th October Anniversary
  Prospect 7A, 117312 Moscow, Russia}
\address[ITEP]{Institute of Theoretical and Experimental physics, B.\
  Cheremushkinskaya 25, 117259 Moscow, Russia}
\address[Juelich]{Forschungszentrum J\"ulich, Institut f\"ur
  Kernphysik, 52425 J\"ulich, Germany}
\address[Seattle]{Department of Physics and INT, University of Washington,
  Seattle, WA 98195-1560, U.S.A.}
\thanks[DFG]{Supported by DFG and RFFI.}
\thanks[hum]{Feodor-Lynen Fellow of the Alexander-von-Humboldt Foundation}

\begin{abstract}
  
  The two-step model with $\pi$, $\rho$ and $\omega$ exchanges taken
  into account is applied to investigate the reactions $p n \to d
  \eta$ and $p n \to d \eta^{\prime}$ near the corresponding
  thresholds.  The existing experimental data on the reaction $p n \to
  d \eta$ are analyzed and predictions for the cross section of the
  reaction $p n \to d \eta^{\prime}$ are presented. It is found that
  $\pi$- as well as $\rho$-meson exchanges yield significant
  contributions in both reactions. Furthermore, the effect of the
  $\eta N$ final-state interaction is studied. It is shown that an
  $\eta N$-scattering length of Re $a_{\eta N} \leq 0.3$ fm is sufficient
  to reproduce the observed enhancement of the $p n \to d \eta$ cross
  section close to threshold.
 

\end{abstract}
\end{frontmatter}

Recently, the total cross section of the reaction $ p n \to d \eta $
has been measured close to threshold at CELSIUS in Uppsala
\cite{Cal1,Cal2}.  These data are remarkable for a variety of reasons:
E. g., they reveal that an older estimate yielding a rather large
near-threshold production cross section for the reaction $ p n \to d
\eta $ \cite{Plouin,Laget} is definitely ruled out \cite{Cal2}.
Furthermore, they show that the production cross section very close to
threshold is significantly enhanced over the expectation for a
two-body phase-space behaviour \cite{Cal2}, indicating the presence of
a strong final-state interaction.

Though there is a large body of theoretical papers about the reaction
$pp\to pp\eta$ only a few of them have also dealt with aspects of the
near-threshold $p n \to d \eta$ and $p n \to (p n)_{I=0}\, \eta$
production \cite{Plouin,Laget,Vet91,Faldt1,Wil98}.  Moreover, all of
them rely on the old and now obsolete data point of
Ref.~\cite{Plouin}. Thus, in the following we want to present a timely
re-analysis of this reaction utilizing now the new data from CELSIUS.
In particular, we are going to present results of a microscopic
calculation of the $p n \to d \eta$ cross section within the framework
of the two-step model (TSM), previously applied to the description of
the Pontecorvo reactions $\bar{p} d \to p M$ (see, e.g.,
Ref.~\cite{Kondratyuk} and references therein). We start from the
$\pi$ exchange contribution to the $\eta$-meson production and study
the relevance of other production mechanisms such as the $\rho$  and
$\omega$ exchange.  Furthermore, we investigate the effect of the
final-state interaction between the $\eta$ meson and the nucleons
employing an optical-potential model as well as a multiple scattering
approach.  Finally, as an application of our model, we present
predictions for the cross section of the reaction $ p n \to d
\eta^{\prime} $ near threshold.

\begin{figure}[htb] 
  \begin{center}
    \leavevmode
    \psfig{file=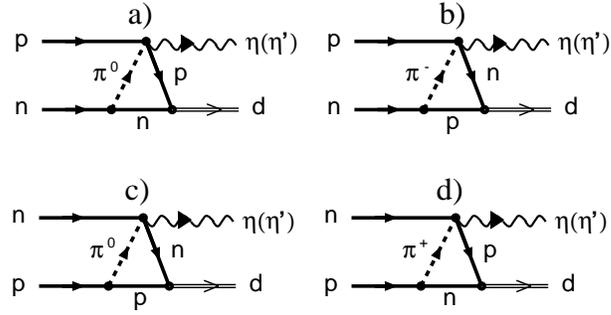,width=8.cm}
    \caption{Diagrams describing the two-step model (TSM). Note that
      besides the $\pi$-exchange contribution also diagrams involving
      the exchange of the $\rho$ and $\omega$ mesons are taken into
      account.}
    \label{fig:tsm} 
  \end{center}
\end{figure}

In the TSM the reaction $p n \to d M $ proceeds in two steps: i) the
initial nucleons produce an $\eta$  or $\eta^\prime$ meson (denoted
as $M$) via virtual $\pi$-, $\rho$- or $\omega$-meson exchange; ii)
the proton and neutron form a deuteron in the final state (see
Fig.~\ref{fig:tsm}).

The amplitude which describes the contribution of the $\pi^0$ exchange
can be written in the form
\begin{equation}
 T_{pn \to dM}^{\pi}(s,t,u)= A_{pn \to dM}^{\pi}(s,t)+ 
A_{pn \to dM}^{\pi}(s,u) \ , 
\label{Atu}
\end{equation}
where $M$ is the pseudoscalar meson $\eta$ or $\eta^\prime$.
$s=(p_1+p_2)^2$, $t=(p_3-p_1)^2$, and $u=(p_3-p_2)^2$ where $p_1$,
$p_2$, $p_3$, and $p_4$ are the 4-momenta of the proton, neutron,
$M$-meson and deuteron, respectively. Treating the nucleons inside the
deuteron nonrelativistically, the two terms on the right hand side of
Eq.~(\ref{Atu}) are given by (see, e.g., Ref.~\cite{KoSa})
\begin{eqnarray}
  A_{pn\to dM}^{\pi} (s,t) &=& \frac{f_{\pi}}{m_{\pi}}
  \varphi^{T}_{\lambda_{2}} ({\vec{p}}_2)\ (-i \sigma_2)
  {\vec{\sigma}}\cdot{\vec{M}}^{\pi}({\vec p}_1)\ {\vec{\sigma}}\cdot
  {\vec{\epsilon}} ^{\ast}_{d}\ \varphi_{\lambda_1} ({\vec{p}}_1)
  \ \times \nonumber \\
  &&  A_{\pi^0 N \to M N}(s_1,t) \ ,
\label{AVdt}
\end{eqnarray}

\begin{eqnarray}
  A_{pn\to d M}^{\pi} (s,u) &=& \frac{f_{\pi}}{m_{\pi}}
  \varphi^{T}_{\lambda_{1}} ({\vec{p}}_1)\ (-i \sigma_2)
  {\vec{\sigma}}\cdot{\vec{M}}^{\pi}(- {\vec p}_1 )\ {\vec{\sigma}}\cdot
  {\vec{\epsilon}} ^{\ast}_{d}\ \varphi_{\lambda_{2}} ({\vec{p}}_2)
  \ \times \nonumber \\
  &&  A_{\pi^0 N \to M N}(s_1,u) \ ,
\label{AVdu}
\end{eqnarray}

where $\vec{\epsilon}_d$ is the polarization vector of the deuteron
and $\varphi_{\lambda}$ the spinors of the nucleons in the initial
state. $f_{\pi}$ is the $\pi NN$ coupling constant, for which we use
the value $f_{\pi}^2/(4\pi ) = 0.08$.  Furthermore we have
\begin{eqnarray}
  &\vec{M}^{\pi}({\vec p}_1 ) &= \sqrt{2\,m} \int ({\vec{k}}+{\vec
    p}_1)\, \Phi_{\pi}({\vec{k}},{\vec{p}_1})\Psi_d(\vec{k})\,
  \frac{\mathrm{d}^3k}{(2\pi)^{3/2}} \ , \label{M} \\
  &\Phi_{\pi}({\vec{k}},{\vec{p_1}}) &=
  \frac{F_{\pi}(q^2)}{q^2-m_{\pi}^2} \ ,
 \label{ff}  
\end{eqnarray}
with the kinematics 
\begin{eqnarray}
  &q^2 &= m_{\pi}^2 - \delta_0(\vec{k}^2+
  \beta(\vec{p}_1)) - 2{\vec {p}}_1 {\vec k} , \
  \vec{q} =  \vec{k}+\vec{p}_1 \ , \nonumber \\
  &\beta(\vec{p}_1) &=
  ({\vec{p}_1}^{\,2}+m_{\pi}^2-T_1^2)/\delta_0\ , \
  \delta_0 = 1+T_1/m, \ T_1= \sqrt{{\vec{p}_1}^2 + m^2} - m \ .
  \nonumber
\end{eqnarray}

In Eqs.~(\ref{M}) and (\ref{ff}) $\Psi_d(\vec{k})$ is the deuteron
wave function, $F_{\pi}(q^2)$ the form factor at the $\pi NN$ vertex
and $m$ the nucleon mass. Note that the amplitude $\vec M^{\pi}$
corresponds to the exchange of a neutral $\pi$ meson only (see the
left diagrams in Fig.~\ref{fig:tsm}). To take into account also the
charged pion exchange we have to multiply amplitude (\ref{Atu})
with an isospin factor (which is equal to 3).

The form factor $F_{\pi}(q^2)$ is taken to be of monopole type with a
cutoff mass $\Lambda_\pi = 0.8 $ GeV/c, in line with recent QCD
lattice calculations \cite{QCD} and other information \cite{Coon}.
The D-wave part of the deuteron wave function gives only a small
contribution and can be savely neglected. This was demonstrated in
Ref.~\cite{Kondratyuk} for the case of the reaction $\bar p d \to M n$
where the same structure integrals (\ref{M}) for $\pi$, $\rho$ and
$\omega$ exchanges occur.  Note that the property
$\vec{M}^{\pi}(-{\vec p}_1) = -\vec{M}^{\pi}({\vec p}_1)$ ensures the
correct behaviour of the amplitude (\ref{Atu}) under permutations of
the initial nucleons. The amplitude should be symmetric in the
isoscalar state.

Near threshold only the S-wave part of the amplitude of the elementary
reaction $\pi N \to M N$ should be of relevance for the production
amplitude $T_{p n \to d M}$.  It can be obtained from the experimental
S-wave cross section through the relation
\begin{equation} |
  A_{\pi^0 N \to M N}(s_1,t)|^2 = | A_{\pi^0 N \to M N}(s_1,u)|^2 = 8
  \pi s_1 \frac{p_{\pi}^{\mathrm{cm}}}{p_M^{\mathrm{cm}}}\sigma_
  {\pi^- p \to M n} 
\end{equation}
where $s_1$ is the square of the c.m.\ energy in the $Mn$ system.  The
values of these S-wave cross sections that we used in our
calculation are \\
$\sigma_{\pi^- p \to \eta n} = (21.2 \pm 1.8) p^M_{\mathrm{cm}}\ \mu$b
and $\sigma_{\pi^- p \to \eta^{\prime}n} = (0.35 \pm 0.03)
p^M_{\mathrm{cm}}\ \mu$b$\\$ ($p^M_{\mathrm{cm}}$ in MeV/c). They are
extracted from the data given in Ref.~\cite{Binnie}.

\begin{figure}[htb]
  \begin{center}
  \leavevmode
  \psfig{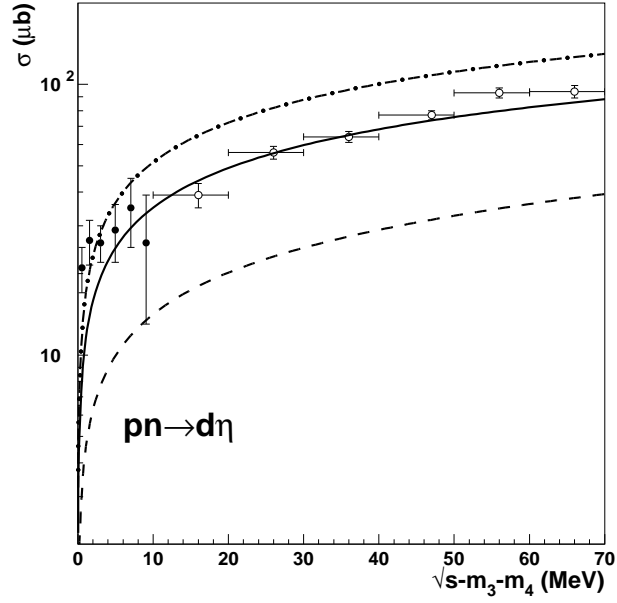}
  \caption{Cross section of the reaction $pn \to d \eta$ 
    as a function of the c.m.\ excess energy.  The dashed curve shows
    the result from the $\pi$-exchange contribution alone whereas the
    dash-dotted curve is the sum of $\pi$, $\rho$, and
    $\omega$ exchange.  The solid curve represents the results
    including all contributions ($\pi$, $\rho$, $\omega$) multiplied
    with a normalization factor $N =0.68$ in order to take into account
    effects from the initial state interaction (see text).  The data
    points for are taken from Refs.~\protect\cite{Cal1} (open
    circles) and \protect\cite{Cal2} (filled circles).}
   \label{fig:etatotal}
  \end{center}
\end{figure}

The total cross sections resulting from the $\pi$-exchange
contribution (with $\Lambda_{\pi} = 0.8$ GeV/c) are shown in
Figs.~\ref{fig:etatotal} and \ref{fig:etaprimetotal} (dashed curves)
as a function of the c.m.\ excess energy $Q = \sqrt{s} - m_3 - m_4$.
The data points for $\eta$ production (Fig.~\ref{fig:etatotal}) are
taken from \cite{Cal1,Cal2}.

Evidently pion exchange alone cannot describe the $\eta$-production
cross-section data. Of course, we could have simply adjusted the
cutoff mass in the pion form factor in such a way that we reproduce
the data. But it goes without saying that the exchange of heavier 
mesons should also be considered as well as effects of the initial- 
and final-state interactions. 

In order to obtain a direct estimate of the contributions from the
$\rho$ and $\omega$ exchanges we calculated them in the framework of
the TSM using the vector-dominance model (VDM) prediction for the
amplitude $\rho N\to \eta(\eta^{\prime}) N$ and assuming that
$A_{\omega N\to\eta(\eta^{\prime}) N} \approx A_{\rho^0 N\to
  \eta(\eta^{\prime}) N}$.  We derive the S-wave $\gamma N\to \eta N$
amplitude from the Mainz data at $E_{\gamma} = 716$~MeV \cite{Mainz}
where the differential cross section of the reaction $\gamma p\to \eta
p$ is isotropic.  In the case of $\eta^{\prime}$ we take the $\gamma p
\to \eta^{\prime}p$ cross section of about 1 -- 1.5 $\mu$b at
$E_{\gamma}$ = 1.5~GeV from Ref.~\cite{Klempt}. According to
Ref.~\cite{Klempt} the differential cross section of the reaction
$\gamma p \to \eta^{\prime} p$ at this energy is almost isotropic. In
the actual evaluation of the amplitudes we use the coupling constants
$\gamma_{\rho}$ and $\gamma_{\omega}$ of Ref.~\cite{Ioffe}. The $V N
\to MN$ amplitudes for longitudinal polarization of the vector mesons
$V$ are taken to be the same as the transversal ones.  Defining the
cross section averaged over the $V$-meson polarizations as
$\mathrm{d}\sigma/\mathrm{d}\Omega$ = $(p_f^{\mathrm{cm}}/
p_i^{\mathrm{cm}}) |f|^2$ we obtain the values $|f|^2$ = 720 $\mu$b/sr
and 60 -- 90 $\mu$b/sr for the reactions $\rho^0 p \to \eta p$ and
$\rho^0 p \to \eta^{\prime} p$, respectively.  Those numbers should be
compared with 360 and 9.8 $\mu$b/sr for the reactions $\pi^0 p \to
\eta p$ and $\pi^0 p \to \eta^{\prime} p$, respectively.

The amplitudes for the vector-meson exchanges can be written in the
form (treating again the nucleons in the deuteron nonrelativistically)
\begin{eqnarray}
A^V_{pn\to dM}(s,t) &=& \frac{G_V}{2\,
  m}\varphi^T_{\lambda_2} (\vec{p}_2) (-i\, \sigma_2) \cdot  A_{V^0N\to
  MN}(s_1,t) \ \times \nonumber\\ 
  &&\left\{ \vec{M}_1^V(\vec{p}_1)\cdot\vec{\epsilon}_d^{\ast} -
         i\left[ \vec{M}_2^V(\vec{p}_1)\times \vec{\epsilon}_d^{\ast}
  \right]\cdot \vec{\sigma} 
  \right\} \varphi_{\lambda_1}(\vec{p}_1) \ , \label{AVt} \\
A^V_{pn\to dM}(s,u) &=& \frac{G_V}{2\,
  m}\varphi^T_{\lambda_1} (\vec{p}_1) (-i\, \sigma_2) \cdot  A_{V^0N\to
  MN}(s_1,u) \ \times \nonumber\\ 
  &&\left\{ \vec{M}_1^V(-\vec{p}_1)\cdot\vec{\epsilon}_d^{\ast} -
         i\left[ \vec{M}_2^V(-\vec{p}_1)\times \vec{\epsilon}_d^{\ast}
  \right]\cdot \vec{\sigma} 
  \right\} \varphi_{\lambda_2}(\vec{p}_2) \ , \label{AVu}
\end{eqnarray}
where
\begin{eqnarray}
\vec{M}_1^V(\vec{p}_1)& =& \sqrt{2\, m} 
  \int [ (\vec{k}-\vec{p}_1) +
  \frac{\vec{k}^2-\vec{p}^2_1}{m^2_V} (\vec{k}+\vec{p}_1)  
  + 2(1+\kappa_V)(\vec{k}+\vec{p}_1)] \ \times \nonumber \\
  &&\Phi_V(\vec{k},\vec{p}_1)
  \Psi_d^{\ast}(\vec{k}) \frac{\mathrm{d}^3k}{(2\pi)^{3/2}} 
\end{eqnarray}
and
\begin{eqnarray}
\vec{M}_2^V(\vec{p}_1) &=& \sqrt{2\, m} 
  \int [ (\vec{k}-\vec{p}_1) +
  \frac{\vec{k}^2-\vec{p}^2_1}{m^2_V} (\vec{k}+\vec{p}_1)] 
  \ \times \nonumber \\
  &&\Phi_V(\vec{k},\vec{p}_1) 
  \Psi_d^{\ast}(\vec{k}) \frac{\mathrm{d}^3k}{(2\pi)^{3/2}} \ .
\end{eqnarray}
$\Phi_V(\vec{k},\vec{p}_1)$ is defined by Eq. (\ref{ff}) where
$m_{\pi}^2$ should be substituted by $m_V^2$ etc. $G_V$ is the
vector coupling constant and $\kappa_V$ is the ratio of the tensor and
vector couplings. For these coupling constants and for the corresponding
vertex form factors we use the values from the full Bonn $NN$
potential \cite{Holinde}, i.e. $G_{\rho}^2/4\pi = 0.84$, $\kappa_\rho = 6.1$,
$G_{\omega}^2/4\pi = 20$, $\kappa_{\omega} = 0$ and 
$\Lambda_\rho$ = 1.4 GeV/c, $\Lambda_\omega$ = 1.5 GeV/c.
Note that the amplitudes (\ref{AVt})--(\ref{AVu}) should also be
multiplied by the isospin factor 3 in the case of $\rho$ exchange. 

As far as the initial state interaction (ISI) is concerned it would be,
of course, desirable to calculate its effect by means of a realistic
model of the $NN$ interaction. However, there are basically no 
$NN$ potentials available in the literature that are still valid in
the energy range where $\eta$ and, in particular, $\eta^\prime$ production
occurs. Therefore we decided to introduce an overall normalization constant
which effectively accounts for the expected reduction of the production
cross sections due to the ISI. This constant is adjusted to the data on
the reaction $pn \to d\eta$ and then used for calculating predictions for
the reaction $pn \to d\eta^\prime$. The latter step is based on the reasonable 
assumption that the ISI does not change that much in the energy
range from the $\eta$- to the $\eta^\prime$-production threshold.

Alternatively, we employ an estimate for the effect of the ISI that
has been presented recently in Ref.~\cite{Hanhart}. It is based on the
assumption that, at the large energies required for the production of
heavier mesons, the $NN$ $t$-matrices are basically constant for
(off-shell) momenta not too far away from the on-shell point.  In this
case the effect of the ISI can be expressed by a simple reduction
factor $\lambda$ that depends only on the $NN$ phase shifts and
inelasticities, i.e.\ \cite{Hanhart}
\begin{equation} 
\lambda(T_{NN}) = \eta_L(T_{NN}) \cos^2{(\delta_L(T_{NN}))} + 1/4\,
[1-\eta_L(T_{NN})]^2 \ .
\label{ISI}
\end{equation} 
In our case the initial $pn$ system is in the $^1P_1$ state. The
corresponding phase shifts and inelasticity parameters near the
thresholds of $\eta$ and $\eta^{\prime}$ production can be taken from
the VA phase-shift analysis \cite{SAID}.  The values at
$T_{\mathrm{lab}} \approx$ 1250 (2400) MeV are $\delta_L(p_1) =
-29^{\circ}(-2.1^{\circ}) $ and $\eta_L(p_1) = 0.54 (0.828)$.  This
yields a reduction factor of $\lambda_{\mathrm{ISI}} =0.46 (0.83)$.

Since the relative phases of the different contributions are not known
we calculate the cross section of the reaction $pn\to dM$ as the
incoherent sum
\begin{equation} 
\sigma_{pn\to dM} = N [\sigma^{(\pi)}+\sigma^{(\rho)}+
\sigma^{(\omega)}] \ .
\end{equation} 

The dash-dotted curve in Fig.~\ref{fig:etatotal} represents the sum
$\sigma^{(\pi)}+\sigma^{(\rho)}+ \sigma^{(\omega)}$ without correcting
for the ISI. This result clearly overestimates the empirical cross
section. In order to achieve agreement with the data for the reaction
$pn \to d\eta$ at $Q > 10$ MeV the theoretical cross section has to be
multiplied by a normalization constant $N$ of 0.68 (see the solid
curves in Fig.~\ref{fig:etatotal}). It is remarkable that this
reduction factor is not very different from the value obtained from
the estimation of ISI effect based on Eq.~(\ref{ISI}). We see that as
a confirmation of the physics contained in our model being basically
correct.

Comparing the dash-dotted with the dashed curve makes clear that we
get significant contributions from the exchanges of vector mesons. We
want to emphasize that those contributions are primarily due to the
$\rho$ exchange. The $\omega$-exchange term is negligibly small and
the result without its contribution would be indistinguishable from
the full result shown in the figure. As a matter of fact, the
magnitude of the $\rho$-exchange term is even larger than the
$\pi$-exchange contribution. This is not unreasonable. Actually,
recently C. Wilkin has argued \cite{Wil98} that a large $\rho$
exchange might be favourable for explaining the observed large cross
section ratio $\sigma_{pn \rightarrow pn\eta} / \sigma_{pp \rightarrow
  pp\eta}$ \cite{Cal3}. We would like to stress that the dominance of
the $\rho$ exchange over $\pi$ exchange in our model is not due to the
soft form factor used for the latter. Even with a value of
$\Lambda_\pi$ = 1.3 GeV as used in the full Bonn model $\rho$ exchange
would still dominate.

The predictions for the reaction $pn\to d\eta^\prime$ are shown in
Fig.~\ref{fig:etaprimetotal}. The dashed curve shows the cross section
when only $\pi$ exchange is taken into account.  The solid and
dash-dotted lines include the contributions from $\rho$ and $\omega$
exchange.  Like in case of $\eta$ production the $\omega$ exchange
gives only a rather small contribution. Thus the difference between
the dash-dotted and the dashed curves is entirely due to the
$\rho$-exchange contribution.  Since there is still a large
uncertainty in the experimental value of the $\gamma p \to
\eta^{\prime} p$ cross section ($1\ \mu$b -- $1.5\ \mu$b)
\cite{Klempt} on which the $\rho$ exchange contribution is based we
present calculations employing the lower and upper bounds.  The
corresponding results (lower and upper solid and dash-dotted curves in
Fig.~\ref{fig:etaprimetotal}) reflect this uncertainty in the input of
our model.  In any case, the $\rho$ exchange increases the cross
section of the reaction $p n \to d \eta^{\prime} $ by almost a factor
of 20--40.  This strong enhancement is caused by the rather large
$\rho N \to \eta^{\prime} N$ S-wave amplitude extracted from the
experimental $\gamma p \to \eta^{\prime}p$ cross section. There are
indications that the latter reaction is dominated by an
$S_{11}(1890)$ resonance, cf.\ Ref.~\cite{Klempt}.
 
Admittedly, the ambiguity involved in the $\rho$ exchange introduce
fairly large variations in the predicted cross section. But still one
can conclude from our model calculation that the cross section for the
reaction $p n \to \eta^\prime d$ for $Q \approx 30$ MeV should be
about $1\ \mu$b or even more, and therefore should be measurable at
new accelerator facilities such as COSY in J\"ulich.

Note that the results shown by the solid curves in
Fig.~\ref{fig:etaprimetotal} were obtained by using the same reduction
factor 0.68 for the ISI as in the reaction $pn\to d\eta$.  Estimate
(\ref{ISI}) suggests a slightly smaller reduction of the $pn\to
d\eta^\prime$ cross section ($\lambda_{\mathrm{ISI}}=0.83$).

In Fig.~\ref{fig:etaprimetotal} we show also the data for the reaction
$pp\to pp\eta^\prime$ which are from \cite{Moskal} (open circles) and
\cite{Hibou} (filled circles) for comparison.  Evidently, near
threshold the cross section predicted for the reaction $pn \to
d\eta^\prime$ is significantly larger than the cross section of the
reaction $p p \to p p \eta^{\prime}$.  This is similar to the case of
$\eta$ production near threshold (see, e.g.\ \cite{Cal1}) and can be
explained by a more favourable isospin factor as well as by the larger
phase space in the reaction $p n \to d M$ as compared with the
reaction $pp \to pp M$.

\begin{figure}[htb] 
  \begin{center}
    \leavevmode
    \psfig{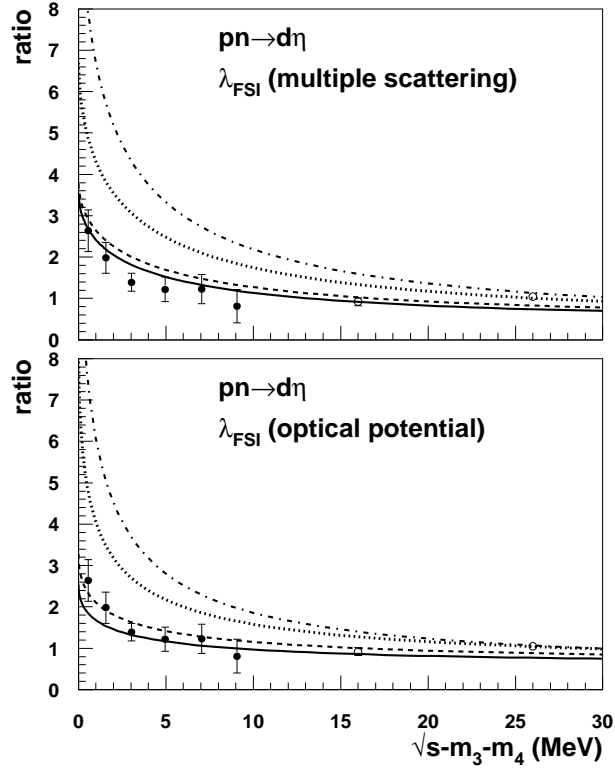}
    \caption{The enhancement factors $\lambda_{\mathrm{FSI}}(p_3)$  
      for different values of the $\eta N$-scattering lengths as a
      function of the c.m.\ excess energy. The solid curves corresponds
      to the $\eta N$ scattering length $a_{\eta N} = 0.291 +
      i\,0.360$ fm derived from $\eta$ photo-production data
      \protect\cite{Krusche}.  The dashed and dotted curves correspond
      to the values $a_{\eta N} = 0.3 + i\,0.3$ fm and $a_{\eta N} =
      0.476 + i\,0.279$ fm proposed in Refs.~\protect\cite{Wilkin} and
      \protect\cite{Faldt2}, respectively.  The dash-dotted curve is
      the result for $a_{\eta N} = 0.717 + i\,0.263$ fm, found in a
      recent analysis of Batini\'c et al.~\protect\cite{Batinic2}.}
    \label{fig:fsi} 
  \end{center}
\end{figure}

It can be seen from Fig.~\ref{fig:etatotal} that the experimental
cross section for very small excess energies lies above the calculated
cross section. In order to demonstrate this more clearly we have
divided the data points by the result of our model (solid curve in
Fig.~\ref{fig:etatotal}) and present this ratio in Fig.~\ref{fig:fsi}.
Then the enhancement of the cross section close to threshold is rather
obvious. It is generally believed that this enhancement of the cross
section at small excess energies is due to a strong and attractive
interaction of the $\eta$ meson with the nucleons in the final state
\cite{Cal2} (a similar enhancement has been also found in the
reaction $pp\to pp\eta$ \cite{Cal96}). Therefore, in the following we
want to discuss the effect of this final-state interaction (FSI) on
the reaction $pn \to d\eta$.

Evidently the enhancement occurs only for $Q \leq 10$ MeV which
corresponds to characteristic momenta of $p_3= p^{\mathrm{cm}}_{\eta}$
below 50 MeV/c. This indicates that the enhancement should be mainly
due to the coherent final-state interaction.  Therefore one can safely
assume that the range of the FSI is much larger than the range of the
interaction that is responsible for the production of the
$\eta$ meson. In this case the (S-wave) production amplitude and the
(S-wave) FSI can be factorised \cite{Goldberger} and the FSI can be
taken into account by multiplying the production cross section
calculated above with a so-called enhancement factor \cite{Goldberger}
resulting from the long range $\eta d$ interaction.

In the simplest version of this approach the enhancement factor
depends only on the $\eta d$ scattering length $A_{\eta d}$ (see,
e.g., Ref.~\cite{Wilkin}).  Values for this scattering length can be
readily found in the literature \cite{Rakityanski,Green}. In the present
investigation, however, we adopt the original prescription of
Goldberger and Watson \cite{Goldberger} where also the $\eta d$ wave
function in the continuum, $\psi_{p_3}(\vec{r})$, is required.
Furthermore, we apply two different approaches for the calculation of
the enhancement factor for the reaction $p n \to d \eta $: The optical
potential (OP) method and the Foldy-Brueckner adiabatic approach based
on the multiple scattering (MS) formalism (see also
Ref.~\cite{Goldberger}).  Note that both these methods have already
been used for the calculation of the enhancement factor for the
reaction $p d \to \,^{3}\!He\ \eta$ (see Refs.~\cite{Wilkin} and
\cite{Faldt2}, respectively).

Let us first describe the OP method.  The optical potential in
momentum space can be expressed through the single $\eta d$-scattering
term
\begin{equation} 
  V_{\mathrm{opt}}(\vec{q}) = -\frac{2\pi}{\mu}\,
  \frac{m_d}{m_{\eta}+m_d}\, [t_{\eta p}(k_{\eta p})+t_{\eta
    n}(k_{\eta n})] \rho_d(\vec{q})\ .
  \label{Vopt}
\end{equation} 
Here $\mu$ is the $\eta d$ reduced mass, and $t_{\eta N}$ the $\eta N$
$t$-matrix which is related to the scattering amplitude $f_{\eta N}$ by
$t_{\eta N}(k_{\eta N}) = (1+\frac{m_{\eta}}{m})\, f_{\eta N}(k_{\eta
  N})$.  Note that we use the scattering length approximation for the
latter, i.e.  $ f_{\eta N}(k_{\eta N})=\left((a_{\eta N})^{-1} -
  ik_{\eta N}\right)^{-1}$.  $k_{\eta N}$ is the modulus of the
relative $\eta N$ momentum and $\rho_d(\vec{q})$ the deuteron form
factor,
\begin{eqnarray}
  \rho_d(\vec{q}) & =& \int\d^3r\,
  \exp{\left(i\frac{\vec{q}}{2}\vec{r}\right)} |\Psi_d(\vec{r})|^2 \ .
\end{eqnarray}
In the $\eta d$ c.m.\ system $\vec{k}_{\eta N}$ is given by 
\begin{equation} 
  \vec{k}_{\eta N} = \frac{m}{m+m_{\eta}}\vec{p}_{\eta} -
  \frac{m_{\eta}}{m+m_{\eta}}\, \frac{\vec{p}_d}{2} =
  \frac{m+m_{\eta}/2}{m+m_{\eta}}\, \vec{p}_3 \ .
\end{equation} 

The optical potential (\ref{Vopt}) is transformed into coordinate
space and then the Schr\"odinger equation is solved numerically
yielding the continuum wave function $\psi_{p_3}(\vec{r}_{\eta d})$ as
well as the $\eta d$-scattering length. The enhancement factor is
simply given by $\lambda^{\mathrm{OP}}_{\mathrm{FSI}}(p_3) =
|\psi_{p_3}(0)|^2$ \cite{Goldberger}.

In the multiple scattering formalism we use the Foldy-Brueckner
adiabatic approximation and start from the $\eta d$ wave function
defined at fixed coordinates of the proton ($\vec{r}_p$) and the
neutron ($\vec{r}_n$) (see Ref.~\cite{Goldberger} for details):
\begin{eqnarray}
  \psi_k(\vec{r}_{\eta},\vec{r}_p,\vec{r}_n) &=& \exp{(i\vec{k} \vec{r}_{\eta})} 
 \nonumber \\
    &+&
    \frac{t_{\eta p}}{D}\, \frac{\exp{(ikr_p)}}{r_p} \,
    \left( \exp{(i\vec{k}\vec{r}_p)} + t_{\eta n}\,
    \frac{\exp{(ikr_{pn})}}{r_{pn}}\exp{(i\vec{k} \vec{r}_n)} \right) \nonumber \\ 
    &+& \frac{t_{\eta n}}{D}\, \frac{\exp{(ikr_n)}}{r_n} \left(
    \exp{(i\vec{k} \vec{r}_n)} + t_{\eta p} \frac{\exp{(ikr_{pn})}}{r_{pn}}
    \exp{(i\vec{k} \vec{r}_p)} \right) \ , \nonumber
\end{eqnarray}
where 
\begin{equation}
  D = \left( 1 - t_{\eta p}t_{\eta n}\,
    \frac{\exp{(2ikr_{pn})}}{r^2_{pn}} \right) \ .
\end{equation}
Here $\vec{r}_{pn}=\vec{r}_p - \vec{r}_n$ and $\vec{k} = \vec{p}_3$ and $k$, $r_{pn}$,
etc., are the moduli of these vectors. $t_{\eta
  N}$ is defined in Eq. (\ref{Vopt}). 
Then the $\eta d$-scattering length and the enhancement factor follow from 
\begin{eqnarray}
  A^{\mathrm{MS}}_{\eta d} &=& \frac{m_d}{m_{\eta}+m_d}\, \langle
  \frac{2 t_{\eta N}(k_{\eta N} =0)}{1-t_{\eta N}(k_{\eta N} =0  )/r}
  \rangle \ , \nonumber \\
  \lambda^{\mathrm{MS}}_{\mathrm{FSI}}(p_3) &=& 
  | \langle \psi_{p_3}(\vec{r}_{\eta}=0,\vec{r}_p=
  \vec{r}/2, \vec{r}_n =- \vec{r}/2) \rangle |^2 \ . \nonumber
\end{eqnarray}
The resulting enhancement factors $\lambda_{\mathrm{FSI}}(p_3)$ are plotted 
in Fig.~\ref{fig:fsi} (upper panel for the multiple scattering
formalism and lower panel for the optical potential) for different values of
the $\eta N$-scattering lengths given in Table 1. In this
Table we also present the calculated values of $A_{\eta d}^{\mathrm{OP}}$
and $A_{\eta d}^{\mathrm{MS}}$ and compare them with the results of
Ref.~\cite{Green}. Of course, the optical potential model is a questionable 
approximation for the $\eta d$ system.
It only gives reasonable values for the $\eta d$ scattering length if
$a_{\eta N}$ is small so that the first iteration is dominant. For larger 
$a_{\eta N}$ it might still give a reasonable estimate of $A_{\eta d}$
provided that there is a cancelation between different higher-order iterations.

Nevertheless, it is expected that the multiple scattering approach
gives a more reliable estimation for $A_{\eta d}$. It is known that
the Foldy-Brueckner approach describes the $\pi d$-scattering
amplitude at low energies quite well (see, e.g.,
Ref.~\cite{Kudryavtsev}). The $\eta$-meson is certainly somewhat
heavier than the pion but its interaction with the nucleons is
likewise considerably weaker than the forces in the $NN$ system.
Therefore the adiabatic approximation might still be quite reasonable
also in the case of $\eta d$ scattering. For example, we see from
Table 1 that within the MS approach an attractive $\eta N $
interaction never yields spurious repulsive results for the $\eta d$
system --- as it happens for the optical potential and for the first
approach used by Green et al.~\cite{Green}.  We want to point out that
our results are in rather good agreement with the second approach of
Ref.~\cite{Green} for a wide range of $a_{\eta N}$ values.

The solid curves in Fig.~\ref{fig:fsi} show our results based on the
$\eta N$ scattering $a_{\eta N} = 0.291 + i\,0.360$ fm derived from
$\eta$ photo-production data \cite{Krusche}. Evidently, they describe
the enhancement of the cross section rather well. The dashed and
dotted curves correspond to the values $a_{\eta N} = 0.3 + i\,0.3$ fm
and $a_{\eta N} = 0.476 + i\,0.279$ fm proposed in Refs.~\cite{Wilkin}
and \cite{Faldt2}, respectively.  The former curve which corresponds
to the smaller value of $a_{\eta N}$ still agrees satisfactorily with
the data, while the latter one overshoots them.  The dash-dotted curve
based on $a_{\eta N} =0.717 + i\,0.263$ fm from a recent analysis of
Batini\'c et al.~\cite{Batinic2} overestimates the data at $Q \leq $
10 MeV by more than a factor 2.  Thus, we conclude that our analysis
of the reaction $p n \to d \eta$ close to threshold favours fairly
small $\eta N$-scattering lengths, Re $a_{\eta N} \leq 0.3$ fm.  The
$\eta d$ scattering length corresponding to these values are $A_{\eta
  d} \approx 0.4 + i\,1.2$ fm, cf. Table 1.  We would like to point
out that the above $\eta N$ scattering lengths are considerably below
the lowest value, Re $a_{\eta N} \approx 0.7$ fm, for which an $\eta
d$ bound state is expected to exist \cite{Green}. Finally, we want to
call attention to the fact that both employed approaches give very
similar results for the enhancement factor.  Furthermore, we would
like to remark that a recent Faddeev-type calculation of the $\eta d$
system testifies that the MS approach is quite reliable for small
values of Re $a_{\eta N}$ like those supported by our analysis
\cite{She98}.

In summary, using the two-step model we calculated the cross sections
of the reactions $ p n \to d \eta $ and $ p n \to d \eta^{\prime} $
near the threshold including $\pi$, $\rho$ as well as $\omega$
exchange. The $\omega$ contribution is found to be rather small. The
contribution of the $\rho$ exchange is of comparable magnitude to the
one from $\pi$ exchange in the case of $\eta$ production and might be
even more important for $\eta^{\prime}$ production due to the presence
of a possible $S_{11}(1980)$ resonance for which the SAPHIR
collaboration found strong evidence in the photo production of
$\eta^{\prime}$ mesons near threshold.  Our results agree with the
experimental data on $\eta$ production if we take into account effects
from the initial state interaction.  We considered also the influence
of the final-state interaction using the optical-potential and
multiple-scattering approaches. Both methods yield similar results for
the enhancement factor. Agreement with the data could be obtained only
for $\eta N$ interactions with small $\eta N$-scattering lengths, Re
$a_{\eta N} \leq 0.3$ fm, i.e.\ values that are considerable below the
lowest value, Re $a_{\eta N} \approx 0.7$ fm, for which an $\eta d$
bound state is expected to exist.

We are grateful to H. Str\"oher for useful discussions.  

\newpage

\begin{table}
\caption{$\eta$-deuteron scattering lengths in fm resulting from the
  optical potential method (OP) and the Foldy-Brueckner approach (MS)
  as described in the text. 'Green1' and 'Green2' refer to results
  presented in Ref.~\protect\cite{Green}.}  \scriptsize
\begin{tabular}{|ll|llll|}
  \hline {} & $a_{\eta N}$ & $A_{\eta d}^{\mathrm{OP}}$ & $A_{\eta d}^{\mathrm{MS}}$
  & $A_{\eta d}^{\mathrm{Green1}}$ & $A_{\eta d}^{\mathrm{Green2}}$ \\
  \hline
  Bennhold-Tanabe \cite{Bennhold} & $0.25+i0.16$ & $0.606+i0.89$ & $0.613+i0.602$ & $0.66+i0.71$
  & $0.66+i0.58$ \\
  Krusche \cite{Krusche} & $0.291+i0.360$ & $-0.14+i1.33$ & $0.399+i1.18$ & $0.17+i1.35$
  & $0.42+i1.25$ \\
  Wilkin \cite{Wilkin} & $0.30+i0.30$ & $0.08+i1.37$ & $0.526+i1.06$ & $0.39+i1.28$
  & $0.58+i1.11$ \\
  F\"aldt \cite{Faldt2} & $0.476+i0.279$ & $-0.15+i2.32$ &
  $1.01+i1.387$ & $0.81+i2.15$ & $1.22+i1.56$ \\ 
  Batini\'c \cite{Batinic1} & $0.888+i0.274$ & $-2.06+i1.37$ & $1.85+i2.83$ & $-2.90+i4.12$
  & $2.37+i5.79$ \\
  Batini\'c \cite{Batinic2} & $0.717+i0.263$ & $-1.94+i2.31$ & $1.62+i2.11$ & $-$  & $-$\\
  \hline
\end{tabular}
\end{table}

\clearpage

\normalsize 

\clearpage

\clearpage

\begin{figure}[htb]
  \begin{center}
  \leavevmode
  \psfig{file=fig2b.epsi,width=8.cm}
  \caption{Cross section of the reaction $pn \to d \eta^\prime$
    as a function of the c.m.\ excess energy.  The dashed curve shows
    the result from the $\pi$-exchange contribution alone whereas the
    dash-dotted curves are the sum of $\pi$, $\rho$, and
    $\omega$ exchange.  The solid curves represent the results
    including all contributions ($\pi$, $\rho$, $\omega$) multiplied
    with a normalization factor $N=0.68$ in order to take into account
    effects from the initial state interaction (see text).  The upper
    and lower solid and dash-dotted curves are the results obtained
    using the maximal and minimal values of the $\rho^0 p \to
    \eta^{\prime} p$ S-wave amplitudes.  The data points are from
    Refs.~\protect\cite{Moskal} (open circles) and
    \protect\cite{Hibou} (filled circles), respectively.}
   \label{fig:etaprimetotal}
  \end{center}
\end{figure}

\clearpage

\begin{thebibliography}{99}

\bibitem{Cal1} H. Cal\'en et al., Phys. Rev. Lett. {\bf 79}, 2642 (1997).
\bibitem{Cal2} H. Cal\'en et al., Phys. Rev. Lett. {\bf 80}, 2069 (1998). 
\bibitem{Plouin} F. Plouin, P. Fleury and C. Wilkin,  Phys. Rev. Lett.
  {\bf 65}, 690 (1990).
\bibitem{Laget} J.-M. Laget, F. Wellers, and J.F. Lecolley,
  Phys. Lett. B {\bf 257}, 254 (1991). 
\bibitem{Vet91} T. Vetter et al., Phys. Lett. B {\bf 263}, 153 (1991).
\bibitem{Faldt1} G. F\"aldt and C. Wilkin, Nucl. Phys. A {\bf 604} 441
  (1996).
\bibitem{Wil98} C. Wilkin, nucl-th/9810047.
\bibitem{Kondratyuk} L.A. Kondratyuk et al., Yad. Fiz. {\bf 61}, 1670 (1998).
\bibitem{KoSa} L.A. Kondratyuk and M.G. Sapozhnikov, Phys. Lett. B {\bf 220},
  333 (1989). 
\bibitem{QCD} K.F. Liu, S.J. Dong, T. Draper, and W. Wilcox,
  Phys. Rev. Lett. {\bf 74}, 2172 (1995). 
\bibitem{Coon} S.A. Coon and M.D. Scadron, Phys. Rev. C {\bf 23}, 1150
  (1981). 
\bibitem{Binnie} D.M. Binnie et al., Phys. Rev. {\bf 8}, 2789 (1973).
\bibitem{Mainz} B.Krusche et al., Z. Phys. A {\bf 351}, 237 (1982);
  Phys. Rev. Lett. {\bf 74}, 3736 (1995). 
\bibitem{Klempt} R. Pl\"otzke et al., Phys. Lett. B {\bf 444}, 555
  (1998). 
\bibitem{Ioffe} V.L.E. Eletsky and B.L. Ioffe, Phys. Rev. Lett. {\bf
    78}, 1010 (1997). 
\bibitem{Holinde} R. Machleidt, K. Holinde, and Ch. Elster,
  Phys. Rep. {\bf 149}, 1 (1987).  
\bibitem{Hanhart} C. Hanhart and K. Nakayama, nucl-th/9809059,
  Phys. Lett. B, in print.  
\bibitem{SAID} R. Arndt et al., Virginia Tech Partial-Wave Analysis Facility, 
  http:// said.phys.vt.edu/
\bibitem{Cal3} H. Cal\'en et al., Phys. Rev. C, in press . 
\bibitem{Moskal} P. Moskal et al., Phys. Rev. Lett. {\bf 80}, 3202 (1998). 
\bibitem{Hibou} F. Hibou et al., Phys. Lett. B {\bf 438}, 41 (1998).
\bibitem{Cal96} H. Cal\'en et al., Phys. Lett. B {\bf 366}, 39 (1996).
\bibitem{Goldberger} M.L. Goldberger and K.M. Watson, {\it Collision theory}
(John Wiley and Sons, New York 1964).
\bibitem{Wilkin} C. Wilkin, Phys. Rev. C {\bf 47} (1993) R938.
\bibitem{Rakityanski} S.A. Rakityanski et al., Phys. Lett. B {\bf
    359}, 33 (1995).  
\bibitem{Green} A.M. Green, J.A. Niskanen and S. Wycech, Phys. Rev. C
  {\bf 54}, 1970 (1996). 
\bibitem{Faldt2} G. F\"aldt and C. Wilkin, Nucl. Phys. {\bf A587}, 769 (1995). 
\bibitem{Bennhold} C. Bennhold and H. Tanabe, Nucl. Phys. {\bf A350}, 625 (1991).
\bibitem{Krusche} B. Krusche, {\it Proceedings II TAPS Workshop}, 
  ed.\ by J. Diaz and Y. Schuts (World Scientific, Singapore 1994,), p.310.
\bibitem{Batinic1} M. Batini\'c et al., Phys. Rev. C {\bf 51}, 2310
  (1995). 
\bibitem{Batinic2} M. Batini\'c et al., nucl-th/9703023.
\bibitem{Kudryavtsev} V.V. Baru and A.E. Kudryavtsev, Sov. J. Nucl. Phys.
        {\bf 60}, 1475 (1997).
\bibitem{She98} N.V. Shevchenko et al., Phys. Rev. C {\bf 58} R3055 (1998).
\end{thebibliography}
\end{document}